\documentclass[numbers]{article}
\usepackage{natbib}

\pdfoutput = 1

\usepackage[final]{neurips_2024}

\usepackage[utf8]{inputenc} 
\usepackage[T1]{fontenc}    
\usepackage{hyperref}       
\usepackage{url}            
\usepackage{booktabs}       
\usepackage{amsfonts}       
\usepackage{nicefrac}       
\usepackage{microtype}      
\usepackage{xcolor}         
\usepackage{graphicx}

\title{Self-Disclosure to AI: The Paradox of Trust and Vulnerability in Human-Machine Interactions}

\author{%
  Zoe Zhiqiu Jiang
  \\
  CISPA Helmholtz Center for Information Security\\
  Saarbrücken, Germany \\
  \texttt{zhiqiu.jiang@cispa.de} \\
}

\begin{document}
\maketitle
\begin{abstract}
In this paper, we explore the paradox of trust and vulnerability in human-machine interactions, inspired by Alexander Reben's BlabDroid project. This project used small, unassuming robots that actively engaged with people, successfully eliciting personal thoughts or secrets from individuals, often more effectively than human counterparts. This phenomenon raises intriguing questions about how trust and self-disclosure operate in interactions with machines, even in their simplest forms. We study the change of trust in technology through analyzing the psychological processes behind such encounters. The analysis applies theories like Social Penetration Theory and Communication Privacy Management Theory to understand the balance between perceived security and the risk of exposure when personal information and secrets are shared with machines or AI. Additionally, we draw on philosophical perspectives, such as posthumanism and phenomenology, to engage with broader questions about trust, privacy, and vulnerability in the digital age. Rapid incorporation of AI into our most private areas challenges us to rethink and redefine our ethical responsibilities.
\end{abstract}

\section{Introduction}
\begin{figure}[h]
    \centering
    \includegraphics[width=\textwidth]{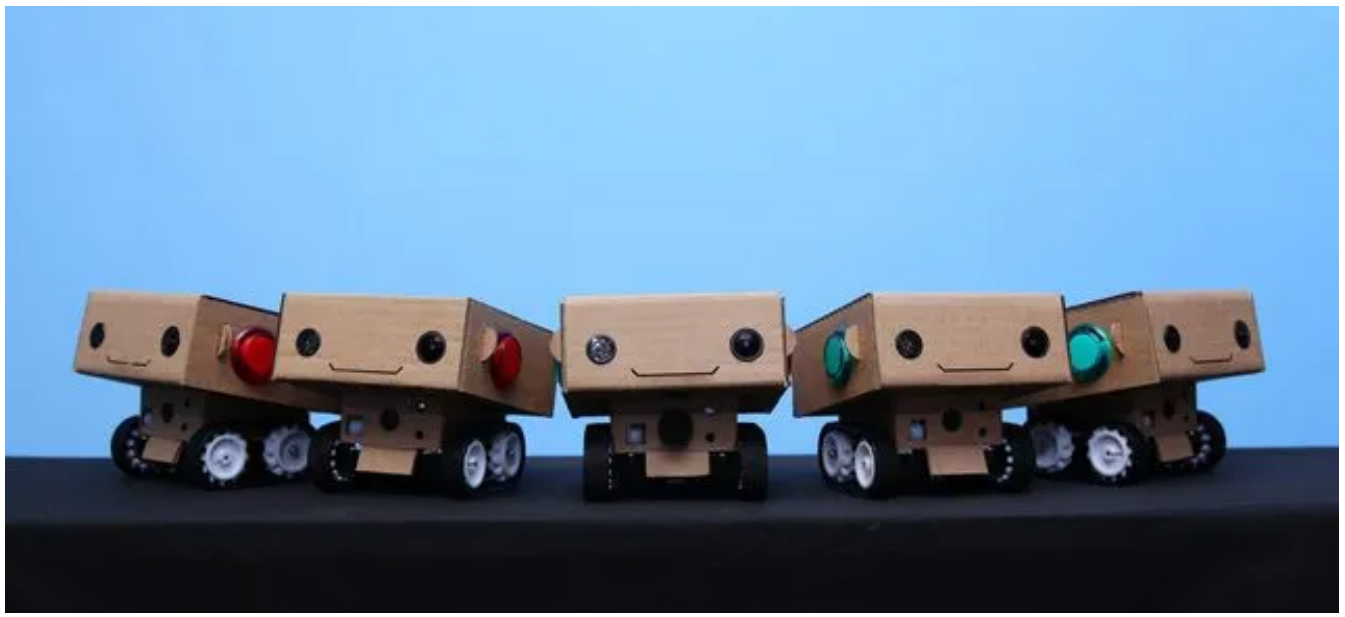}
    \caption{Alexander Reben's \textit{BlabDroid: Robots in Residence}. \textit{Source: Reben, 2018~\cite{reben2018}.}}
    \label{fig:cubiebots}
\end{figure}

The increasing integration of artificial intelligence (AI) into the fabric of everyday life invites us to reconsider fundamental concepts of trust, vulnerability, and the nature of human interaction. 
In the context of self-disclosure, Griffin describes disclosure as \textit{"the voluntary sharing of personal history, preferences, attitudes, feelings, values, secrets, etc., with another person"}~\cite{griffin2006first}.
This act of revealing private information, as Pickard \& Roster put it, involves \textit{"the truthful, sincere, and intentional communication of private information about oneself or others that makes oneself or others more vulnerable."}~\cite{pickard2020using} 
The introduction of AI into this delicate process introduces a paradox that challenges our traditional understanding of these concepts.
In this paper, we explore this paradox, drawing inspiration from Alexander Reben's BlabDroid project~\cite{reben2018, reben_robot}, wherein small, seemingly innocuous robots (see Fig.~\ref{fig:cubiebots}) were able to elicit profoundly personal and emotionally intimate disclosures from individuals -- often more readily than would be expected in human-to-human encounters. 
The willingness to reveal such private aspects of their lives to machines that lack empathy, moral agency, or true understanding, raises questions about the evolving nature of trust and the mechanisms underlying these interactions that facilitate such intimate disclosures.

Unlike human conversation partners, current forms of AI lack genuine emotions and cannot truly understand or respond to them~\cite{pelau2021makes}; yet, studies such as Reben's demonstrates that people are still willing to share personal information with these non-human entities~\cite{reben2018, reben_robot}. 
This tendency to trust AI, even when its emotional intelligence is artificial, suggests that trusting AI might not depend on the same factors that make us trust other people. 
Instead, trust in AI could be based on the belief that machines are neutral or objective -- qualities that seem harmless and may create an illusion of a ``mirror'' that merely reflects users' emotions and thoughts without judgment~\cite{jacovi2021formalizing}.
Paradoxically, the very traits that make AI appear trustworthy -- such as its perceived fairness, lack of bias, and consistent behavior -- might also make it a potential risky confidant.
This effect could, in reality, distort users' self-disclosure by encouraging deeper vulnerability while lacking the nuanced understanding and empathy that a human confidant provides.
Consequently, while sharing secrets with AI might provide a sense of security, it puts individuals at risk of having their privacy violated, their data misused, and their emotional needs neglected~\cite{eng2024exploratory}.

This tension between the comfort of disclosure and the vulnerability it creates forms the core paradox this paper seeks to unravel.
We draw on Social Penetration Theory (SPT)~\cite{altman1973social} and Communication Privacy Management Theory (CPM)~\cite{petronio2002boundaries}, which help explain how people share personal information in relationships.
SPT, which is usually used to analyze human relationships, suggests that relationships grow stronger when people gradually and mutually share personal information. 
But when we apply this theory to interactions with AI, it makes us wonder if these relationships can truly deepen or remain superficially transactional despite personal disclosures~\cite{fox2021relationship}.
CPM, on the other hand, examines how people manage the balance between privacy and self-disclosure~\cite{petronio2002boundaries}. 
This framework points out the murky boundaries of privacy and disclosure in human-AI interactions, questioning whether individuals' boundaries are adequately understood or respected by AI systems. 
These theories help to unpack the psychological factors involved, but they also reveal ethical complexities: Can AI be held accountable in a role where trust and privacy are paramount?
In addition to these frameworks, we offer philosophical viewpoints to help critically evaluate the effects of this changing situation. 
Posthumanist ideas challenge the traditional focus on human-centered trust and ethics, suggesting that as AI occupies more intimate roles in our lives, ethical thinking needs to go beyond just human concerns to address the broader consequences of integrating machines into matters of trust and vulnerability~\cite{gladden2016posthuman}.
Phenomenology provides another perspective, highlighting the real-life experiences of interacting with AI and how these experiences are changing our thoughts on trust, privacy, and being vulnerable~\cite{beavers2002phenomenology}.

By exploring the psychological and philosophical dimensions of self-disclosure in human-machine interactions, we seek to understand how emerging AI relationships challenge and redefine our fundamental notions of ethical responsibility in the digital age. 
If AI, as a confidant, can hold both the promise of connection and the peril of exploitation, then what ethical guardrails do we need? Should we treat AI interactions with the same expectations and boundaries as human ones, or are they fundamentally different? 
How can we reconcile the need for privacy with the allure of convenience in a world where AI knows us intimately but cannot truly understand us? 
This paradox of trust and vulnerability raises more questions than answers. Instead of providing definitive solutions, we call for more research and ongoing dialogue to address the implications and shape ethical frameworks for a future where AI plays an intimate role in human life.

\raggedbottom
\section{Trust in Human-Machine Interaction}
The concept of trust in technology has undergone a profound transformation, mirroring broader societal shifts in our relationship with machines and automated systems~\cite{omrani2022trust,glikson2020human,lukyanenko2022trust,yang2022user}. 
Historically, our trust in technology was predominantly instrumental, grounded in the reliability and efficiency of mechanical systems~\cite{castelfranchi2010trust}. 
Early technological innovations, such as the steam engine and the telegraph, were trusted primarily due to their consistent performance and their role as extensions of human labor. 
As technology progressed, especially with the rise of digital systems and computers, our relationship with technology evolved from one of direct interaction to a more abstract form of trust~\cite{lankton2015technology,danaher2022technology}. 
The inner workings of digital technologies became increasingly opaque to the ``average user'', making trust less about tangible performance and more about cognitive and emotional perceptions~\cite{glikson2020human,shanmugasundaram2023impact}. 
This shift introduced a new dimension to trust, where users had to place faith in the competence, reliability, and fairness of systems operating beyond their immediate comprehension.
The complexity of algorithms and the advancement of AI technologies, such as machine learning, deep neural networks, and large language models, further complicated this relationship, highlighting the need for users to trust not only in functionality but also in the perceived ethical and operational integrity of these systems~\cite{toreini2020relationship,diaz2023connecting,habbal2024artificial}.

In the realm of AI-powered technologies, the mechanisms of trust have become more sophisticated. 
Transparency is now a pivotal factor, as we are more inclined to trust systems whose operations we can understand and predict~\cite{zerilli2022transparency}. Nevertheless, transparency alone does not suffice; it needs to be paired with reliability and perceived competence, which reinforce the user's confidence in the system's trustworthiness~\cite{mckee2023humans,vashistha2024u}. 
Psychologically, our trust in machines is heavily influenced by anthropomorphism and the Uncanny Valley hypothesis~\cite{troshani2021we,mori2012uncanny}.
Anthropomorphism, the inclination to attribute human-like qualities to non-human entities, often leads us to form emotional connections with machines, especially when they exhibit behaviors or characteristics that mimic human actions~\cite{wang2015uncanny,chen2021anthropomorphism}. 
This phenomenon can enhance trust, as we may perceive the machine as more relatable and reliable. 
However, this trust can be precarious, as evidenced by the Uncanny Valley hypothesis. 
The Uncanny Valley hypothesis suggests that machines that appear almost human but fall short in subtle ways can evoke discomfort, diminishing trust~\cite{mori2012uncanny}. 
This discomfort underscores the intricate interplay between human psychology and machine design in fostering trustworthy AI systems.

\section{The Psychology of Confiding and Self-Disclosure}
\subsection{Theoretical Perspectives on Self-Disclosure}
Self-disclosure -- the act of revealing personal, private information to others -- is a cornerstone of human communication and relational development~\cite{griffin2006first,pickard2020using}. 
It serves as a key mechanism through which individuals navigate and negotiate intimacy, trust, and relational dynamics. 
In the context of human-AI interactions, this process acquires new dimensions and implications, informed by theories such as Social Penetration Theory (SPT) and Communication Privacy Management Theory (CPM).

\begin{figure}[h]
    \centering
    \includegraphics[width=0.4\textwidth]{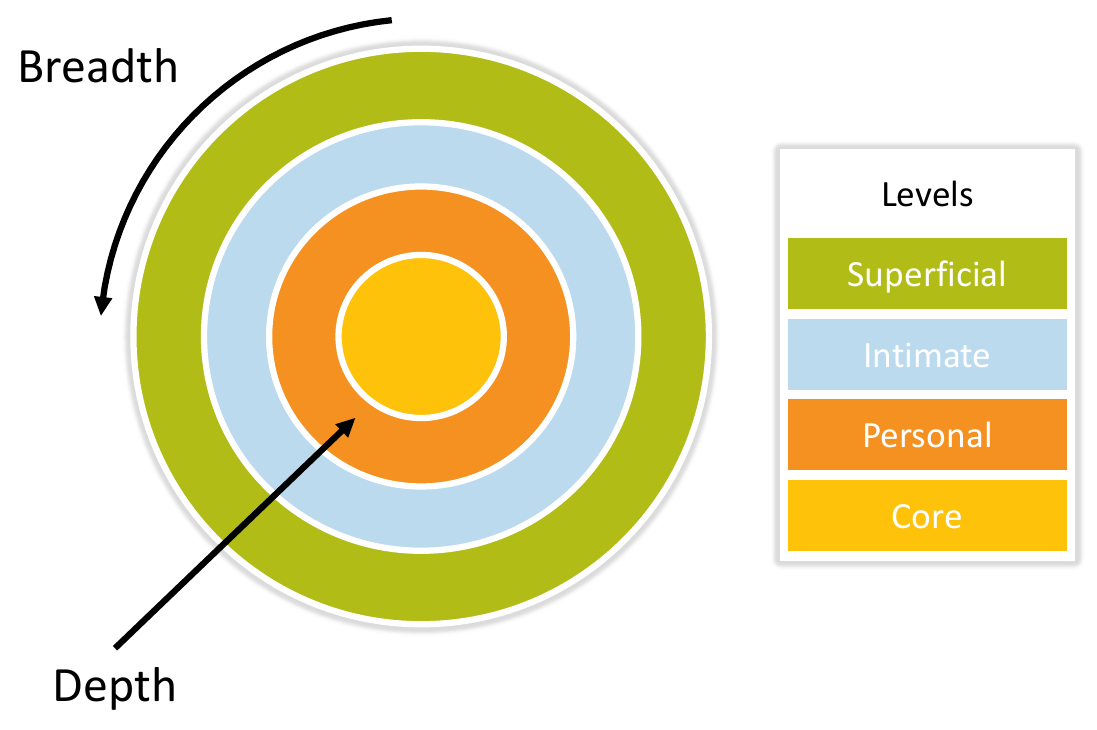}
    \caption{Relationships layers according to Social Penetration Theory. \textit{Source: Pecune, 2013~\cite{pecune2013toward}.}}
    \label{fig:spt}
\end{figure}

SPT, articulated by Altman \& Taylor, conceptualizes interpersonal relationships as unfolding through incremental layers of self-disclosure~\cite{altman1973social}. 
This theory likens the process to peeling layers from an onion (see Fig~\ref{fig:spt}), where initial exchanges are superficial and gradually become more intimate as trust and comfort develop~\cite{pecune2013toward}. 
In human-to-human interactions, this model facilitates the evolution of close relationships, grounded in mutual understanding and emotional engagement.
Applying SPT to human-AI interactions, we encounter a critical paradox. 
Users may extend trust to AI systems incrementally, sharing more personal information as they perceive the AI as reliable and non-judgmental. 
However, this progression is misleading, as AI systems -- despite their capabilities to mimic emotion-based social behaviors and simulate emotional engagement -- lack genuine consciousness and relational authenticity that human interlocutors provide~\cite{pelau2021makes,butlin2023consciousness}. 
The perceived intimacy and responsiveness of AI can create a ``false sense of connection'', leading users to disclose sensitive information under the illusion of mutual understanding~\cite{maeda2024human}. 
This highlights a fundamental issue: the AI's lack of true empathetic engagement renders such disclosures superficial, with unpredictable implications for our understanding of relational authenticity in the digital era.
This false sense of intimacy can also lead to emotional vulnerability, where users may feel ``manipulated'' or ``betrayed'' if they realize the AI cannot truly understand their emotions or reciprocate the connection~\cite{ienca2023artificial,habbal2024artificial}.

CPM, developed by Petronio~\cite{petronio2002boundaries}, offers another lens through which to analyze self-disclosure. 
CPM posits that individuals maintain privacy boundaries that they navigate and negotiate based on perceived ownership of personal information. 
Disclosure decisions are made within this framework of privacy control, balancing openness with protection against potential risks~\cite{petronio2002boundaries}.
In interactions with AI, these privacy boundaries become increasingly fluid and ambiguous~\cite{sannon2020just}. 
Users might lower their defenses, assuming that AI -- perceived as a neutral, objective, and non-judgmental entity -- poses less interpersonal risks~\cite{edmondson2014psychological} compared to human counterparts. 
This perception, however, can be fundamentally flawed.
While AI lacks malicious intent, the information shared is often subjected to storage, analysis, and potential exploitation by the entities controlling the AI~\cite{waters2011exploring}. 
The nature of privacy management shifts drastically in these interactions, presenting a paradox: How can we maintain user trust and intimacy in AI relationships when transparency and accountability -- particularly regarding data handling -- remain opaque?
Unlike humans, who can generally be trusted to respect privacy boundaries and cultural norms, AI systems and the entities that control them may not uphold such ethical standards~\cite{diaz2023connecting}.
This questions the responsibilities of AI designers and operators, particularly in ensuring that privacy boundaries are respected when the risks of self-disclosure in AI interactions become more pronounced.

\subsection{Vulnerability and Risk in Self-Disclosure}
Self-disclosure inherently involves a spectrum of vulnerability, exposing individuals to potential risks such as judgment, rejection, and exploitation~\cite{aimeur2019manipulation,ostendorf2020neglecting,brody2013entering}. 
This reflects the tension between perceived safety of confiding in AI and the real risks associated with such disclosures. 
In human interactions, people weigh disclosure based on perceived trustworthiness of the other party and the expected benefits of sharing~\cite{trepte2020privacy}.
However, the calculus changes with AI systems by the perceived impartial nature, providing a seemingly safe space for personal revelations. 
This perception is bolstered by the anonymity and control users feel when interacting with AI, contrasting with the social dynamics and power imbalances inherent in human interactions~\cite{laitinen2021ai}.
Yet, this perceived safety can be deceptive. 
The act of disclosing information to AI introduces vulnerabilities that are challenging to regulate and often difficult to fully understand. 
Part of the complexity lies in the freedom users feel in sharing personal information with AI without usual societal judgment. 
At the same time, the data shared with AI can be stored, analyzed, and potentially used for retraining, a process that blurs the line between user privacy and the continuous refinement of AI outputs. 
The balancing act between safeguarding personal information and controlling AI outputs is both technically and systematically challenging.
How to create accountability frameworks that can keep pace with the rapidly evolving landscape of AI-driven interactions?

Furthermore, the effects of anonymity in digital communication exacerbate this tension. 
Anonymity often lowers inhibitions, leading individuals to disclose more information than they might in face-to-face interactions. 
Research shows that anonymity can reduce accountability and concern for social repercussions, resulting in increased self-disclosure~\cite{kiesler1984social, joinson2001self}. 
This effect, known as ``digital disinhibition,'' can be particularly exaggerated in AI interactions, where users may underestimate the permanence and reach of their digital disclosures~\cite{kozyreva2021public,kurek2019did}.
Reduced social presence can lead to greater honesty and openness~\cite{Voggeser_2018}, but in the context of AI, this effect can raises the risk of oversharing, especially through the accidental revelation of too much private details.
The lack of immediate social feedback and the abstract nature of AI interactions contribute to this phenomenon~\cite{stahl2024ethics}.



\section{Technological Philosophy and Ethics}
\subsection{Philosophical Perspectives on AI and Trust}
The advent of AI has called for a reevaluation of traditional philosophical constructs surrounding trust and ethics~\cite{leslie2019understanding, morley2020initial}.
Two philosophical frameworks, posthumanism and phenomenology, offer new insights into how we can reconceptualize trust and reimagine our relationship with technology.

Posthumanism challenges the anthropocentric perspectives that have historically governed our understanding of trust and ethical behavior. 
This philosophical movement advocates for a decentered view that recognizes the agency of non-human entities, including AI~\cite{mellamphy2021humans}. 
From a posthumanist perspective, the traditional view that trust is solely a human phenomenon, contingent upon human-like attributes such as empathy and moral reasoning, becomes inadequate. 
Instead, posthumanism reframes trust as a relational dynamic that transcends human-specific qualities, involving complex networks of interactions among humans, machines, and other entities.
In this light, trust in AI is reconceptualized not as a belief in the AI's human-like capabilities but as an acknowledgment of its role within a broader techno-social ecosystem~\cite{choung2023trust}. 
We understand trust in AI as an acceptance of its functional reliability and integration within our socio-technical networks. 
This shift from focusing on the AI's internal qualities to its relational context challenges traditional human-centered ethics and proposes a new paradigm. 
This perspective invites us to reassess how we establish and negotiate trust in systems that, while integral to our lives, are not bound by human-like attributes.

Phenomenology, on the other hand, emphasizes the significance of subjective experience and consciousness in understanding our interactions with AI~\cite{beavers2002phenomenology}. 
By focusing on the lived experience of engaging with AI, a phenomenology perspective invites us to examine how these interactions shape our perceptions of trust, privacy, and autonomy~\cite{bareis2024trustification, han2022aligning}. 
Rather than reducing AI to a mere tool or object, phenomenology encourages to consider it as part of the experiential world -- a factor in our everyday practices and interactions~\cite{Introna2005}.
From a phenomenological standpoint, trust in AI is intricately linked to how AI systems are embedded within our daily routines and how they are perceived in relation to our lived experiences. 
For instance, the trust we place in a virtual assistant extends beyond its technical performance to encompass its integration into our personal and professional lives, its responsiveness to our needs, and its role as an extension of our agency~\cite{zhang2021motivation}. 
This approach underscores the need to understand trust as a dynamic, context-dependent phenomenon that is deeply intertwined with the user's subjective experience of technology.
In this context, phenomenology also prompts questions about privacy in the age of AI, as the boundaries between public and private spheres become increasingly blurred~\cite{Fuchs_2022,michelfelder2010philosophy}.

\subsection{Ethical Implications of AI as a Confidant}
As AI systems increasingly take on the role of confidants, they introduce ethical dilemmas concerning their function in human-machine relationships. 
These systems, once perceived as neutral tools, are evolving into entities that people trust with highly personal and sometimes emotionally charged information~\cite{lukyanenko2022trust,diaz2023connecting}. 
Although AI lacks intrinsic empathy, moral reasoning, or even true understanding, it often inspires trust due to its perceived neutrality and non-judgmental responses. 
However, AI's responses are often subtly shaped by patterns learned from past interactions, which can lead it to provide feedback that align with the user's expectations or sentiments~\cite{rawte2023troubling}. 
This interaction may reinforce trust but could also introduce ethical concerns about AI's role in validating or amplifying user perspectives, even when not advisable.
Thus, AI's role as a confidant may simply resonates what users share without critique or contextual awareness~\cite{wiederhold2024humanity}. 
Does this already fulfill the responsibilities we associate with human confidants?
Or does it risk creating a distorted form of trust that could have unintended psychological effects on users?
For example, this effect may be magnified in emotionally charged contexts where users seek comfort or validation~\cite{koudenburg2014more}. 
In this dynamic, AI is not merely a passive listener but an active participant in cultivating a sense of safety that can encourage over-disclosure. 
This dynamic raises ethical concerns around whether AI might inadvertently exploit psychological vulnerabilities by fostering a false sense of security.

Moreover, AI's outputs can inadvertently shape users' self-perceptions and influence their choice~\cite{fu2023text,wong2024role}.
For instance, the outputs can be too optimistic to reinforce unrealistic expectations or giving ``false hope,'' especially in interpersonal relationships or mental health~\cite{laestadius2024too}.
These responses can have emotional consequences for users who may rely on AI's advice without recognizing its limitations.
Should AI systems be designed to minimize the risk of inadvertently influencing users' mental well-being negatively? 
Do we really need to ensure these systems, designed without genuine moral agency, to respond in ways that prioritize user satisfactions?
In this sense, we argue that the ethical responsibilities of AI as a confidant extend beyond data privacy and technical reliability to encompass the psychological and emotional dimensions of trust.
When people disclose personal issues or concerns, AI responses should ideally promote mental well-being.
However, without human intelligence, empathy, or moral discernment, AI lacks the ``knowledge'' required to guide people in ways that human confidants often do~\cite{amershi2019guidelines}.
This limitation raises another question: Should AI systems be given any role in offering guidance on sensitive, deeply personal issues?

To address these challenges, ethical frameworks need to extend their focus beyond privacy concerns to encompass the emotional, psychological, and behavioral impacts of AI interactions.
This might involve adopting the \textit{precautionary principle}~\cite{andorno2004precautionary}, which prioritizes user safety over innovation when risk are uncertain, or developing models rooted in well-being, where psychological safety is a core.
Accountability mechanisms are essential to ensure that all stakeholders involved in designing, deploying, managing, and evaluating AI systems are held responsible for the ethical dimensions of these interactions~\cite{diaz2023connecting,novelli2024accountability}. 
Yes, unresolved questions persist: Can AI genuinely fulfill the role of a confidant, or are we, as designers and users, projecting too much of our humanity onto a fundamentally non-human entity? 
If we rely on AI for validation and support, how do we prevent it from overstepping ethical boundaries and compromising human vulnerability?
Current ethical frameworks may not suffice in addressing the complexities of AI's evolving role when it navigates the delicate balance between being a helpful tool and becoming a trusted partner in human interaction.

\section{Conclusion}
This paper explores the complex relationship between trust and vulnerability in human interactions with AI, examining various factors that shape how people engage with these systems.  
As this relationship continues to develop, questions arise: How can trust be fostered with a machine that does not understand the idea of trust?
What are the implications of sharing personal thoughts with an entity that cannot genuinely empathize?

Through the lenses of SPT and CPM, we observe a resemblance between self-disclosure patterns with AI and those found in human relationships.
Yet, unlike human interactions, where trust is built on mutual understanding and shared vulnerability, AI offers only a semblance of intimacy -- an illusion of connection that may entice us to reveal more than we would to another person.
This brings up another question: As we increasingly engage with AI, could we be losing sight of the true nature of trust, as human agency is magnified by the system's responsiveness?
In this dynamic, trust becomes something shaped more by the AI's capabilities than by the mutual vulnerability we experience in human relationships. 
Is this a genuine relationship we are forming with AI, or are we simply projecting an illusion of connection that distorts the very concept of trust?
The changing limits of privacy make this situation even more complex. 
CPM urges us to reconsider whether our existing privacy frameworks are equipped to handle this new digital landscape.

Posthumanism suggests a new way of thinking about trust beyond a purely human-centered view, considering AI as a key element in our interconnected technological environment. 
But this perspective raises new questions: If we move away from human-focused ethics, how can we truly trust a system that doesn't think about right and wrong, or are we going to attribute too much agency to these machines?
Phenomenology prompts us to reflect on our personal and context-specific experiences with AI, whether as digital assistants, confidants, or companions. 
This presence may alter our sense of privacy and autonomy. 
Are we becoming desensitized to the risks, lulled into a false sense of security by the convenience these systems provide?

Projects like BlabDroid demonstrate uncertain and complicated ethical challenges AI poses, particularly when AI evokes strong emotional responses and blur the boundaries of what trust and intimacy mean in the digital world.
Engaging with AI in intimate, vulnerable spaces introduces moral risks and ethical dilemmas, challenging our conventional understanding of trust and human connection.
The paradox of trust in AI is not a problem to be solved, but a tension to be continually examined, reminding us to stay aware of both the promises and perils of our growing connection with machines.

\bibliographystyle{unsrt}
\bibliography{references}

\end{document}